# Flexible ferromagnetic nanowires with ultralow magnetostriction

Giuseppe Muscas[1,3], Petra E. Jönsson[1], I. G. Serrano[1], Örjan Vallin[2], and M. Venkata Kamalakar[1]*

[1]Department of Physics and Astronomy, Uppsala University, Box 516, SE-751 20 Uppsala, Sweden
[2]Department of Engineering Sciences, Uppsala University, Box 534, SE-751 21 Uppsala, Sweden
[3]Department of Physics, University of Cagliari, Cittadella Universitaria di Monserrato, S.P. 8 Km 0.700 (Monserrato-Sestu), I-09042 Monserrato CA, Italy

*Email: venkata.mutta@physics.uu.se

**Abstract:** *Integration of magneto-electric and spintronic sensors presents a massive potential for advancing flexible and wearable technology. Magnetic nanowires are core components for building such devices, and therefore it important to realize flexible magnetic nanowires and uncover magneto-elastic properties, which can propel not only such flexible sensing applications, but can also make new pathways for exploration of flexible magneto-plasmonic devices, and discovering unseen observations at reduced dimensions. Here, we realize ferromagnetic nanowires on flexible substrates for the first time. Through extensive magneto-optical Kerr experiments, exploring the Villari effect in such nanowires, we reveal a two-order of magnitude reduced magnetostrictive constant in nanowires, compared to bulk values. In addition, the nanowires exhibit a remarkably resilient behavior sustaining bending radii ~ 5 mm, very high endurance, and enhanced elastic limit compared to thin films of similar thickness and composition. We confirm the observed performance by micro-magnetic simulations and attribute the observations to the size reduction and high nanostructure-interfacial effects. The flexible magnetic nanowires with ultralow magnetostriction open up new opportunities for stable surface mountable and wearable spintronic sensors, enable a credible way for engineering advanced nanospintronic devices and exploring new effects in hybrid heterostructures.*

**Keywords:** flexible nanoelectronics, nanowires, magneto-elastic coupling, spintronics, magnetostriction

Advancement in flexible electronics is key to next-generation flexible sensors, wearable devices, and micro-nano medical applications. In the arena of such emerging developments, the integration of magnetic and spintronic components into flexible electronics presents a multitude of new prospects. Furthermore, flexible magnetic nanowires are prime ingredients for the recent realization of flexible spin circuits[1], as well as advancing magnetic sensors[2], spin-wave engineered magnonic[3], and novel flexible magneto-plasmonic components. This makes it fundamental to realize flexible nanomagnets and determine their magneto-elastic limits. In contrast to conventionally investigated magnetic thin films and heterostructures[4], magnetic nanowires on flexible substrates have never been studied, primarily due to several challenges related to their fabrication, unknown adhesion and clamping effects to flexible substrates, as well as the challenges in precisely capturing their magnetic response to strain. The phenomena of inverse magnetostriction, also called as the Villari effect, where altering the size of a magnetic material by applying stress results in modified magnetic characteristics, needs to investigated here. The magnetostrictive behavior in bulk ferromagnetic metals such as Fe, Co, and Ni is well known and arises from spin-orbit coupling[5]. The effect can be simply pictured as an externally applied magnetic field, forcing the alignment of spins resulting in a distortion of electronic clouds, which induces a variation in interatomic distances. The link of exchange interaction with interatomic distance or the magneto-elastic coupling makes magnetism quite sensitive to strain. Due to the finite size effects, nanowires exhibit unique scaling and phase transition behavior[6–9], which also makes magnetostriction in ferromagnetic nanowires and their impacts non-trivial. Therefore, realizing and establishing the magneto-elastic limits of flexible magnetic nanowires will make a significant impact fundamentally as well as technologically.

In this work, we realize highly stable ferromagnetic nanowires on flexible polyethylene naphthalate (PEN), an emerging low roughness substrate (average RMS roughness ~ 1.3-1.5 nm) for flexible electronic applications. By optimizing the adhesion and fabrication conditions, we achieve a large-dense-array of non-interacting magnetic nanowires (as shown in Fig. 1) with exceptional resilience upon bending. Through magneto-optic Kerr effect (MOKE) experiments, we comprehensively probe magnetism and Villari effect in ferromagnetic nanowires fabricated on flexible substrates subjected to bending radii up to 5 mm to unlock magneto-elastic coupling. By performing micromagnetic simulations, we confirm and assess the magnetoelastic properties of the nanowires. These experiments and simulations reveal, for the first time, the remarkable resilient nature of the magnetic nanowires, with high yield strength, and ultralow magnetostriction.

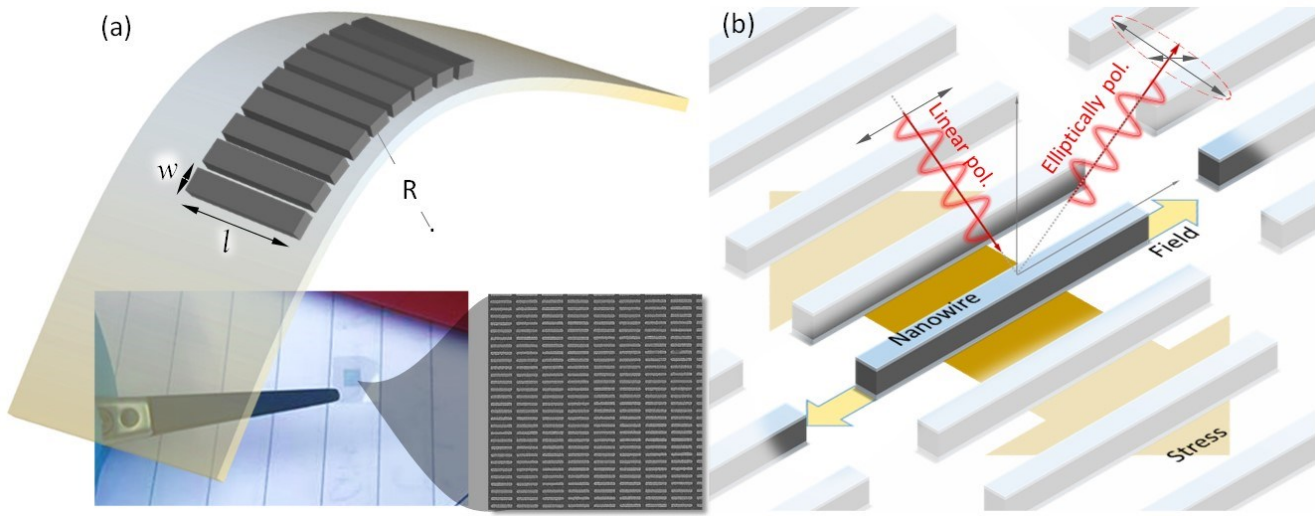

Figure 1. (a) Sample scheme of flexible nanowires along with bottom panel showing 4 million nanowire (200 nm x 2 µm) arrays and an SEM image showing a large array of nanowires. (b) Scheme of Magneto-optic Kerr Effect employed to probe flexible magnetic nanowires.

For this work, we fabricated arrays of nanowires on flexible PEN substrates by optimized e-beam lithography, e-beam evaporation, and metal liftoff processing. The final nanowire array displayed in Figure 1a consists of a layered structure of 3 nm Ti |30 nm Co| 4nm Al stack, with good clear separation of nanowires over large area ~ 4 mm x 4 mm. In comparison to fabricating thin films, obtaining such large nanowires areas without metal residues is not very straight forward. However, utilizing these conditions, we obtained such large arrays for nanowires for both 100 nm and 200 nm wide nanowires with up to 30 nm of Co thickness. Considering the substrate roughness, we observed that an initial layer of 3 nm Ti promotes uniform metal adhesion throughout the surface. In addition to these nanowire samples, for reference, we also prepared a flexible cobalt thin film having the same-layered structure. A key challenge of this study was to investigate the properties of nanowire samples mimicking real-world conditions of flexible devices. Magnetic nanowires exhibit sharp magnetization reversal process, i.e., the switching between the two antiparallel magnetic saturation states at a precise coercive field ($H_C$). Hence, the study of the magnetic properties of such a small nanostructure requires a high sensitivity, which needs to be optimized for measurements under controlled amount of stress. To capture the magnetization reversal process for both nanowire and thin-film samples, we recorded magnetization versus field M(H) loops developing a home-built longitudinal magneto-optic Kerr effect (L-MOKE) setup as depicted in Figure 2a MOKE scheme. To bend these samples stably, we fabricated special 3D printed curved sample holders with several radii of curvatures, which provide stable support to the samples while allowing us to control the curvature precisely. Shown in the lower panel of Fig. 2a, the 3D printed sample stage for sample mounting consisted of a curved mold with a square window cap to clamp the flexible substrate to the curve base (see Methods).

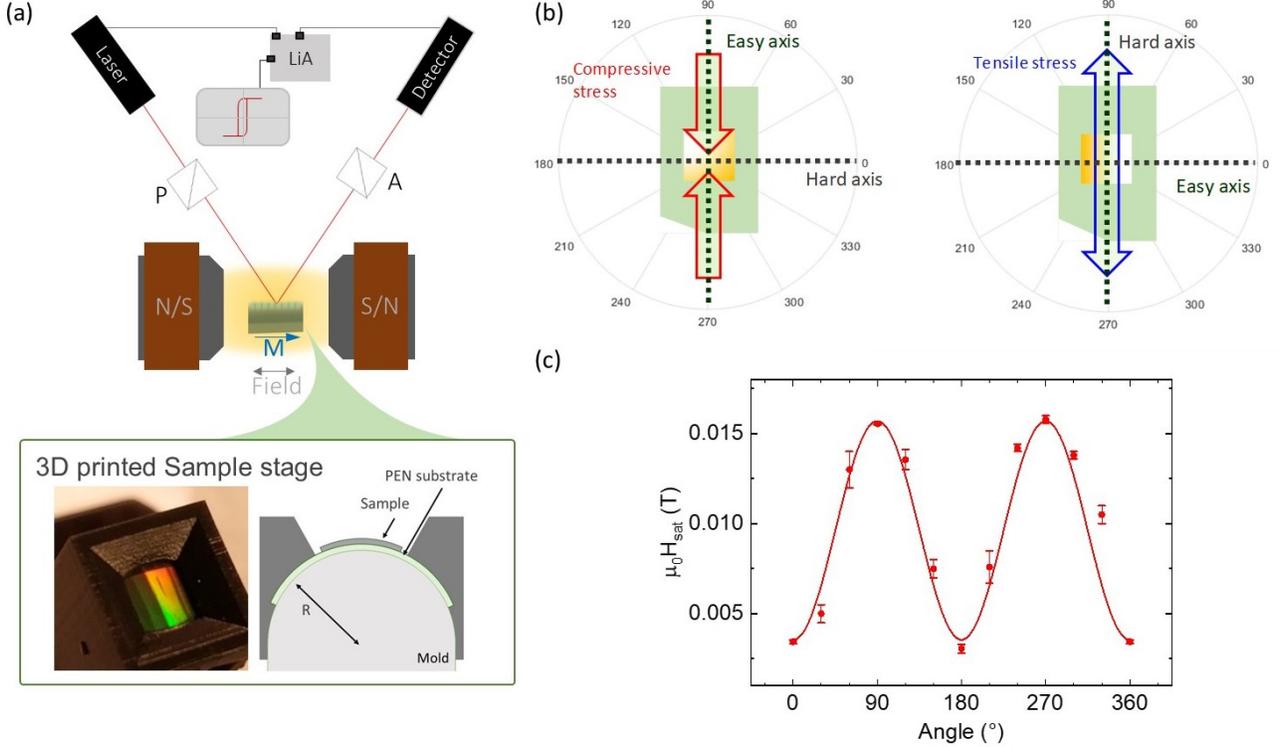

**Figure 2.** (a) Top panel: scheme of magnetization measurements on nanowires (thin films) using longitudinal magneto-optic Kerr effect (L-MOKE) along with experimental arrangement consisting of laser and polarizer (P) for the emission of linearly polarized light, followed by the analyzer (A) and detector arrangement involving a Lock-in amplifier (LiA) for measuring the reflected elliptically polarized light. Bottom panel (left to right): scanning electron microscope image showing an array of nanowires, a 3D printed sample stage with a bent sample of a thin film, and a cross-sectional view of the stable curved sample stub. (b) Representation of how tensile stress and compressive stress induce hard and easy magnetic anisotropy axes in ferromagnets with negative magnetostriction such as cobalt [10]. (c) Saturation field as a function of the applied filed angle. The fit to the data allows estimating a residual stress anisotropy of about 8500(300) J m$^{-3}$.

For a specific radius of curvature of the sample holder (R), the induced strain ε on the sample is given by[11]:

$$\varepsilon = \frac{h+t}{2R+h+t} \qquad (1)$$

where, $h$ is the thickness of the flexible PEN substrate, and $t$ is the magnetic layer thickness. The corresponding stress $\sigma$ exerted on the material, with Young modulus $E$ and Poisson's ratio $\gamma$ of the material, can be calculated using Eq. (2) given below [12]

$$\sigma = \frac{\varepsilon E}{(1-\gamma^2)} \qquad (2)$$

Such stress induces a uniaxial magnetic anisotropy component with corresponding anisotropy energy $E_\sigma$ [5]:

$$E_\sigma = K_\sigma sin^2\theta = \frac{3}{2}\lambda_\sigma \sigma sin^2\theta \tag{3}$$

where, $K_\sigma$ is the stress-induced anisotropy constant, $\lambda_\sigma$ the magnetostrictive constant of the material, and $\theta$ is the angle between the magnetization and easy axis, with $\lambda_\sigma$ = -60 × 10$^{-6}$ for bulk polycrystalline cobalt[10]. The negative sign of $\lambda_\sigma$ implies that the application of tensile stress promotes an anisotropy easy axis perpendicular to the applied stress, while compressive stress produces an easy axis parallel to it, as illustrated in Fig. 1b.

First, we probe the Co films by measuring M(H) curves after rotating the sample at different in-plane angles with respect to the applied field. Due to the polycrystalline nature of the sample, the random distribution of the easy axis of each grain should, in principle, average out the anisotropy, leading to a negligible effective value with no angular dependence[13]. However, a weak but clear uniaxial anisotropy emerged, as revealed by the angular dependence of the saturation field extracted from M(H) curves (see Figure 2b), which indicates the high sensitivity of our setup to characterize magnetism in thin films faithfully. Note that the observed anisotropy is normal for magnetic thin films occurring due to intrinsic stress developed during the preparation process of thin films[14]. By bending the substrate, we then applied different values of tensile and compressive stress perpendicular to the intrinsic easy axis, while measuring the M(H) loops with the field applied parallel (Figure 3a) and perpendicular to the film's intrinsic easy axis (Figure 3b). The figures are consistent with the negative magnetostriction of cobalt, for which an external tensile (compressive) stress produces an additional anisotropy with easy axis perpendicular (parallel) to the stress direction. Consequently, such an additional stress-induced anisotropy can add to or compete against the intrinsic anisotropy resulting in an increase or decrease of the original coercivity, in accordance with the geometrical configuration, as illustrated in Figure 3a-b. Here, combining Eq. 1 and Eq. 2, we have estimated the external stress applied to the film as a function of the radius of curvature. For calculations, we have used Young's modulus E = 209 GPa reported for nanocrystalline Co[15] and a Poisson's ratio γ = 0.31 of bulk Co[16], considering that both theoretical and experimental results on γ show no appreciable deviation from the bulk value for nanostructured metallic alloys[17–19]. This is supported by the fact that within the classical finite size effect regime, metal nanowires of the present dimensions exhibit characteristic Debye temperature[6,8] of the same order as bulk values.

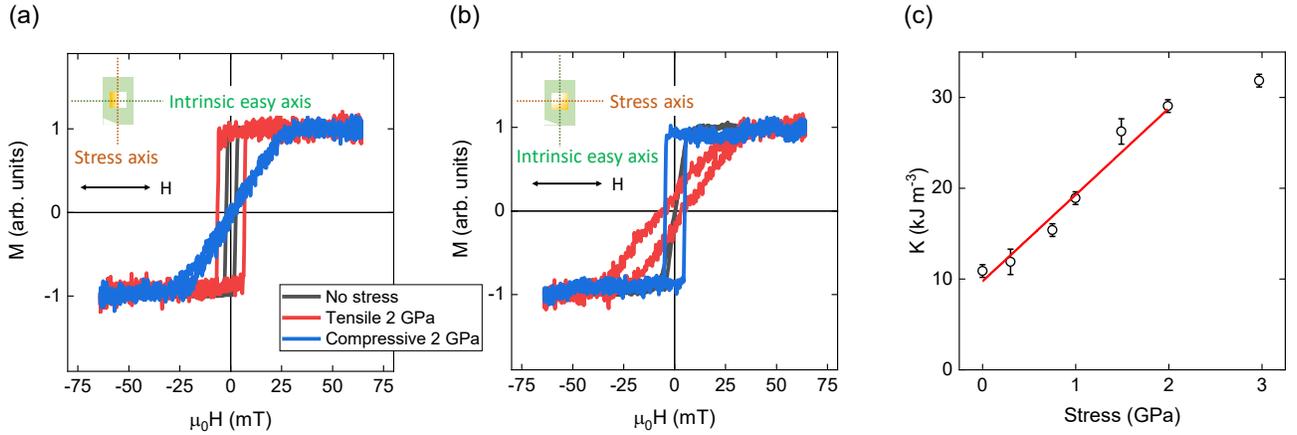

**Figure 3.** M(H) curves of the Co film before bending (black curves), after a 2 GPa tensile stress (red curves), and a 2GPa compressive stress (blue curves) applied perpendicularly to the intrinsic easy axis of the film. The curves are measured with the magnetic field parallel (a) and perpendicular (b) to the original easy axis, as depicted in insets. (c) Values of the effective anisotropy of the film under different tensile stress applied (calculated from Eq. 4). The red line shows the fit to the data within the elastic regime of the film.

Using the saturation field $\mu_0 H_{sat}$ along the hard axis induced by the applied tensile stress, we can estimate an effective anisotropy as:

$$K = \frac{\mu_0 H_{sat} M_S}{2} \tag{4}$$

assuming the bulk saturation magnetization $M_S$ = 1.4 MA/m[10]. In **Figure 3c**, we present the experimental values of anisotropy as a function of in-situ applied tensile stress. The stress was applied perpendicular to the intrinsic easy axis, as shown in the inset of **Figure 3a**. A modified eq. 3, $K_{eff} = K_{intr} + \frac{3}{2}\lambda_s\sigma$, that encompasses the intrinsic anisotropy of the film, was fitted to the $K_{eff}$ values within the elastic (linear) regime (Figure 3c). The fit reveals a magnetostriction constant $\lambda_\sigma$

= - 6.3(5) × 10$^{-6}$, which is one order of magnitude smaller than the value reported for bulk polycrystalline cobalt[20]. Such a reduction is consistent with similar behavior observed previously in FeGa films[21]. The fitting procedure was limited to the practical elastic regime of the sample, i.e., up to 2 GPa of external stress. Note that the film thickness is within the range for efficient stress transfer[22,23], and we did not notice any physical change or thin films to be detached from the substrates during or after the experiments. Beyond physical observations, the most sensitive way to detect permanent plastic deformation is offered by the magnetoelastic behavior itself, where we see a marked deviation from the expected linear elastic regime, only for 3 GPa of external stress. Therefore, in general, our method has the additional value of probing with high precision the mechanical properties of nanostructures, allowing the observation of the most extended elastic regime so far reported for Co thin films.

After confirming the mechanical stability and adhesion of our thin-film heterostructures on PEN, we focused on measuring the magnetic response of nanowires (NWs) fabricated with the same layered structure on PEN substrates. Cobalt nanowires with a width of 200 nm and a length of 2 µm are organized in arrays (Fig. 1a), with a horizontal and vertical spacing pitch as low as 500 nm to minimize magnetic interactions between the individual nanowires. In contrast to continuous thin films, the shape anisotropy of the nanowires establishes a strong anisotropy easy axis along its length. Applying the magnetic field parallel to the length of the nanowires, we recorded M(H) loops under different values of applied stress. The configuration of the applied stress, magnetic field, and nanowire geometry, along with the scheme of measurements on nanowires, are displayed in **Figure 1b**. For the nanowires, tensile and compressive stress is applied perpendicularly to the shape anisotropy easy axis. A comparison of the M(H) loops under initial flat conditions, as well as the loops measured with an applied tensile and compressive stress of ~2 GPa, is displayed in **Figure 4a**. A summary of the extracted coercivities for different applied stress is presented in **Figure 4b**. The observed trend of coercivity enhancement with tensile stress and reduction with compressive stress is in tune with the fact that tensile stress strengthens the anisotropy along the intrinsic easy axis, increasing the coercivity and the compressive stress acts oppositely. With the large shape anisotropy of the nanowires, the magnetization reversal can be reasonably described by the Wolfarth-Stoner model[24]. Therefore, for an M (H) loop measured along the easy axis of an individual single nanowire, we can attribute its behavior to that of a single magnetic domain structure, where $H_c$ ~ $H_{sat}$, hence we can rearrange Eq. 3 and 4:

$$H_C = H_C^0 + \frac{3}{M_S}\lambda_s\sigma \quad (5)$$

where $H_C^0$ is the original intrinsic coercivity. Using $M_S$ = 1.4 MA/m, valid also in case of cobalt nanowires[25,26], we fit $H_c$ versus stress using eq. 5 to estimate the magnetostriction in the case of tensile stress yielding the anisotropy easy axis along the long axis of the nanowire. Strikingly, tensile stress reveals an ultra-small $\lambda_\sigma$ = -9 (2) × 10$^{-7}$, which is nearly two orders of magnitude smaller than that of bulk Co[15]. Therefore, the magnetic behavior of nanowires remains mostly unaltered despite the application of stress.

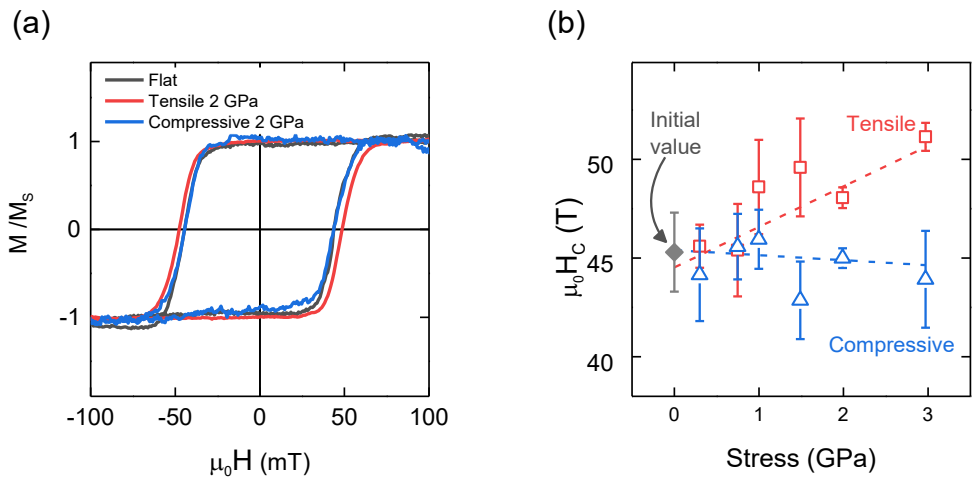

**Figure 4.** (a) M(H) loops measured along the length (easy axis) of the Co nanowires, in flat and applied stress conditions. (b) Coercivity for the NWs extracted from M(H) loops as a function of in-situ applied stress along with linear fits to Eq. (5).

Although simple, this model allows describing a clear picture of the magneto-elastic properties of our nanowires from the evolution of coercivity, a key parameter for the application of magnetic nanowires in the context of flexible-devices. The present nanowires are above the conventional size for a coherent reversal of the magnetization[10,24]; hence one cannot exclude a priori eventual domain wall formation and motion and its role on the observed coercivity. Furthermore, while the external stress induces a corresponding stress anisotropy over the entire NW, locally, they are constituted by Co nanograins with intrinsic magnetocrystalline anisotropy and randomly oriented easy axis. This randomization averages out the overall effective magnetocrystalline anisotropy[27]; nevertheless, it is essential to identify its eventual influence over the magnetization reversal. Finally, one could argue that the presence of a dominating shape anisotropy hinders any other effect of external stress. The complexity of the overall scenario has never been systematically investigated, and therefore, to find answers to these questions, we have created a micromagnetic model of the nanowires using MuMax$^3$ [28], a GPU accelerated micromagnetic simulation software (Details in methods).

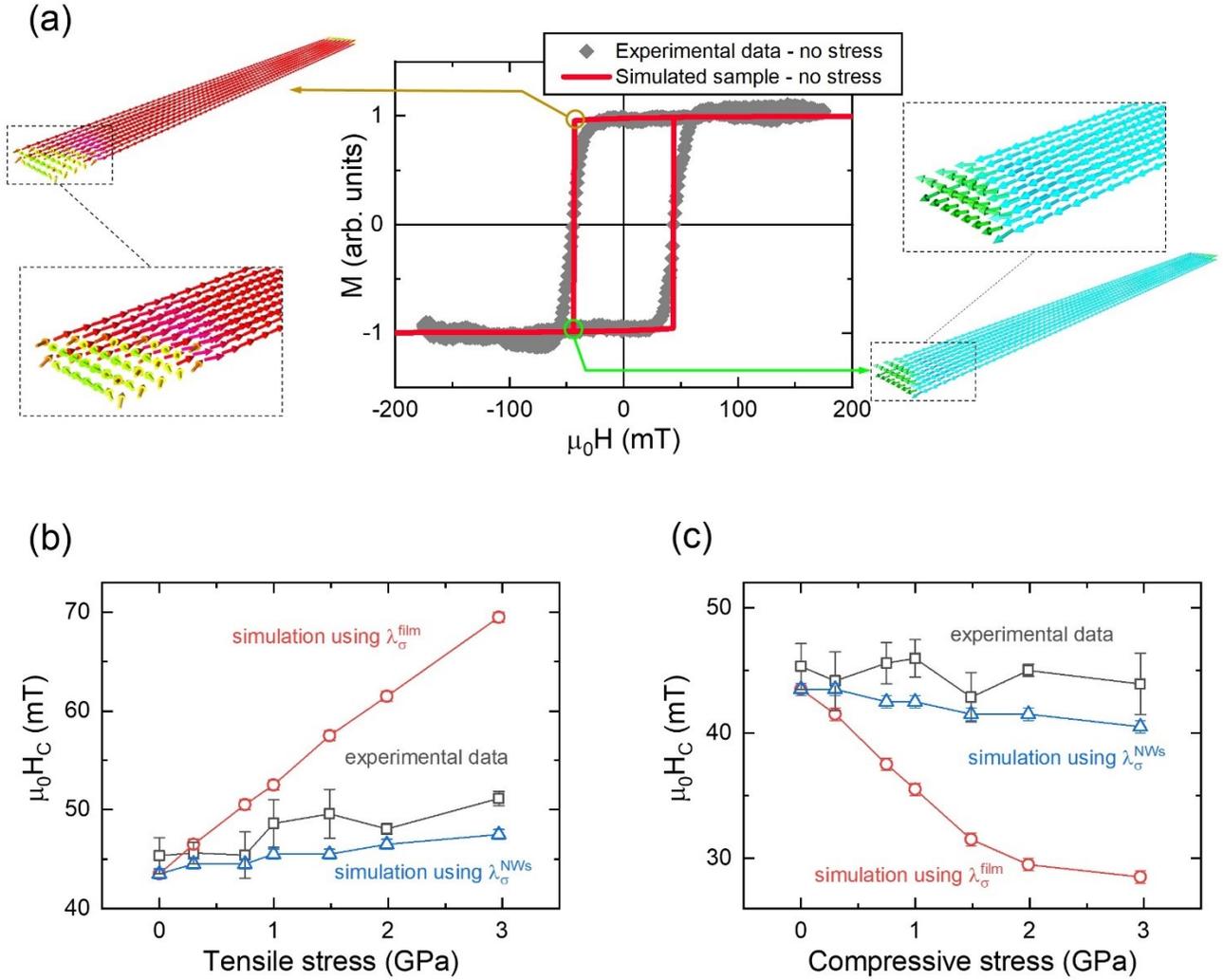

**Figure 5.** (a) Experimental M(H) curve and corresponding simulated one for an un-stressed nanowire, along with details of the curled magnetic configuration during magnetization reversals. Comparison of the simulated stress-induced variation of the coercivity of nanowires using the experimentally obtained $\lambda_\sigma$ for nanowires and thin films under (b) tensile stress and (c) compressive stress.

As shown in Fig. 5a, we obtained an excellent qualitative and quantitative description of the experimental data by using typical values of saturation magnetization ($M_S$ = 1.4 MA/m) and adjusting the exchange stiffness to take into account an effective room temperature exchange across the nanocrystalline structure ($A_{ex}$ = 17.5 pJ/m)[29,30]. In addition, we have tested the role of magnetocrystalline anisotropy repeating the simulation with and without a random anisotropy model (RAM) including a grainy structure (10 nm grains) with typical intrinsic magnetocrystalline anisotropy (K = 520 kJ/m$^3$)[31] and random distribution of easy axis. Using or not RAM leads to the same results for both stressed and unstressed systems. This confirms that the intrinsic magnetocrystalline anisotropy has a negligible influence, due to the random direction of easy axis among grains and the dominant effect of the shape anisotropy. The second important observation is that the NWs magnetization never forms domain walls, but just minimal curling at the edges at fields close to the magnetization reversal, which happens in a coherent-like mode, under the driving force of the shape anisotropy, as evident by the insets of **Figure 5a**. This reinforces the significance of our basic model used to treat the experimental data. To test the effect of external stress we have implemented its effective uniaxial anisotropy in the model. First, the magnitude of the stress anisotropy has been calculated using the values of external applied stress and the effective magnetostriction constant $\lambda_\sigma^{film}$ obtained from the Co thin film samples. The coercivity of the NW varies according to the trend presented in **Figure 5b-c**. It is evident that if the nanowires would own the same magnetostriction of the thin film, their variation of coercivity will be not quenched by intrinsic anisotropy magnetocrystalline or by the shape anisotropy and the increment of $H_c$ would be much larger than what experimentally observed. Repeating the same simulations using the experimental effective magnetostriction constant $\lambda_\sigma^{NWs}$ of the NWs, we could reproduce the experimental $H_c$ vs stress trend (**Figure 5b-c**). Hence, these results strongly reinforce the thesis of an ultrasmall effective magnetostriction constant in the NWs, much smaller than that of the thin film.

The implementation of magnetic nanowires in real-world flexible device applications demands robust mechanical stability. In order to test the resilience of our nanowires on the flexible substrates, we performed bending of the wires 100 times with a radius of curvature of 5.0 mm, which corresponds to an applied tensile stress of ~3 GPa. As evident in **Figure 6a**, the hysteresis behavior of the nanowires is faithfully restored, even after 100 bending cycles. We observed similar robust behavior in nanowires with lower width (See Figure S1). To check for deviations accrued during bending, we set the samples back to the flat configuration after each cycle of bending stress and performed measurements in the flat configuration. The result of such bending tests for nanowires as compared to the Co thin films of similar thickness is displayed in Figure 6b. As evident here, for thin films, stress applied in different directions up to a value of 2 GPa did not induce any permanent deformation, while stress ~3 GPa permanently increases the original coercivity (nearly 40% increase), which is a signature of the plastic deformation of the sample. This is because, after the plastic deformation, when the substrate is brought back to its original flat configuration, the elongated sample retains tensile stress, which leads to the observed higher coercivity. From these results, we can conclude that the yield strength, the maximum stress that a film can sustain before permanent plastic deformation, surpasses 2 GPa, which is a significantly high value compared to bulk Co and it is in tune with studies performed on thicker nanocrystalline films[15]. The normalized $H_c$ versus subjected stress plot reveals that nanowires can sustain higher stress than thin films of the same thickness on PEN, which indicates an enhanced elastic limit of at least 3 GPa in the case of nanowires. These experiments enable us to conclusively demonstrate that the cobalt nanowires are highly resilient and endure an extreme bending radius ~ 5 mm (~1.4% strain) of the flexible substrate, which is of fundamental significance to developing stable flexible spintronic and magneto-electric sensors.

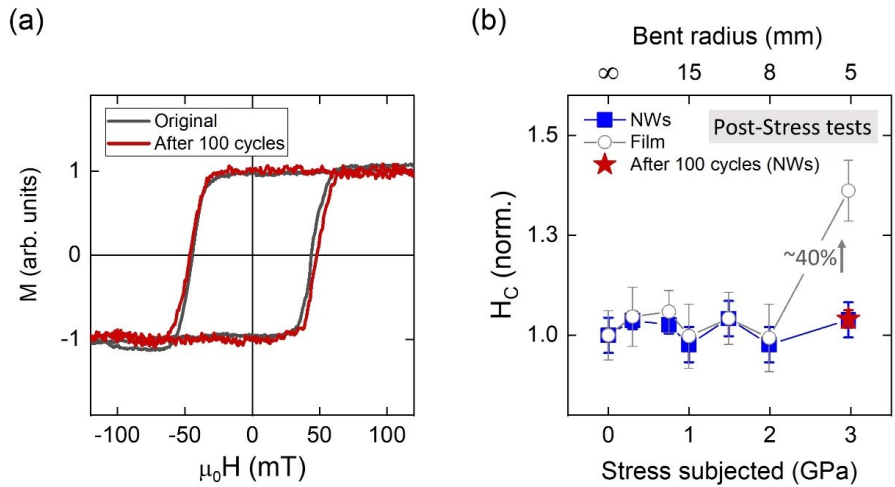

**Figure 6.** (a) Comparison between the original (black line) M(H) hysteresis of 2.0 μm × 0.2 μm NWs and after bending them 100 times (red curve) applying 3 GPa of tensile stress (radius of curvature of 5.0 mm). (b) Post-stress tests showing normalized coercivity of cobalt nanowires and thin film: the normalized coercitivity is the ratio of the coercivity measured in the flat configuration after each bending (subjected to different stress) to the starting value.

High resilience of the cobalt nanowire demonstrated is non-trivial, but feasible considering the fact that improved elastic properties can be anticipated for nanostructured metals[15,32]. Artificially structured magnets display unusual magnetic properties, due to the unique atomic environment at the surface/interface, characterized by missing bonding or bonding between different species. The high concentration of topological defects at the grain boundaries and surface can lead to magnetostrictive behavior contrasting bulk magnetostriction[20,33,34]. Specifically, localized magnetostrictive strains at the interfaces are responsible for surface magnetostriction contributions[35]. For magnetic thin films, the surface magnetostriction competes with the bulk value, leading to the linear dependence of the effective magnetostriction constant on the inverse thickness in accordance with Szymczak's empirical model[35]. The model has been successfully employed to characterize the behavior of thin films for Fe and Co-based alloys[36,37]. A generalization to the Szymczak model[35] for nanowires can be made as $\lambda_{eff} = \lambda_{bulk} + \lambda_{surface} (S/V)$, where an enhanced surface to volume ratio (S/V) could bring down the effective magnetostriction $\lambda_{eff}$ as observed in our experiments. Additional contributions to the deviation with respect to bulk can emerge from the eventual strain at the substrate interface[34], the surface roughness of the system[38], and possible crystalline structural reorganization in thin layer[22]. This implies that the further reduction of the effective magnetostriction to the reduced dimensionality of the NWs from 2D to 1D geometry is due to enhanced S/V and the enhanced local topological modifications and nanostructured interfaces, making the surface magnetostriction properties dominating over the bulk behavior. For each step of reduced dimensionality (bulk -> film -> NWs), nearly one order reduction in the magnitude of $\lambda_\sigma$ has been observed[3,6–9,39]. Hence, our experiments demonstrate, for the first time, ultralow effective magnetostriction resulting in the coercivity of nanowires remains nearly unaltered up to large bending radii of curvature (~5 mm). The high elastic limit (at least 3 GPa), ultralow magnetostriction, and the stability of the magnetic properties after 100 cycles of bending suggest a highly resilient magneto-elastic coupling in nanowires fabricated on flexible PEN.

For a broad spectrum of applications, the quenched magnetostriction forms the basis for developing stable magnetic sensors, spintronics, and novel magneto-plasmonic application that can be integrated to advance flexible electronics and wearable sensing technologies. Flexible magnetic nanowires with high magneto-elastic resilience with no mechanical fatigue is of principal advantage to devices, where one does not desire any change in magnetic properties and device performance due to the bending of magnetic elements but aims for extreme stability, for instance, in surface mounted sensors, steady magnetic switches, and spin valves. Over the past decade, the field spintronics has seen rapid progress in the direction of planar multi-terminal spin circuits, made possible due to the advent of graphene and other graphene-like two dimensional (2D) crystals[40–42]. In metal or graphene planar spin devices, ferromagnetic nanowires are widely employed to generate and detect spin current. The recent flexible spin circuits of graphene[1], opened a new door to utilize the exceptional resilience of 2D materials for flexible 2D spintronics. Here, the current work provides crucial advancement in achieving magnetically and mechanically stable ferromagnetic nanowires as electrodes, which is key for harnessing strain-induced band structure quantum changes [12] in 2D materials spin valves for novel sensing and transduction. The resilience of ferromagnets can also enable unveiling the impact of strain on novel spinterfaces[42,43] for strain-induced tuning of spin polarization in spintronic devices, exploration of strain-controlled magneto-coulomb single electron spintronic transducers[44]. In addition, flexible magnetic nanowires are prospective candidates for developing tunable spin-wave engineered magnonic devices[3], new magneto-thermopower[45] applications, while the transparency provides ways to manipulate light via innovative magneto-plasmonic devices. Hence, the flexible magnetic nanowires enable new possibilities for multiple kinds of flexible and mechano-magnetic applications.

**Conclusion**
In summary, we have demonstrated flexible magneto-elastic nanowires with exceptional flexibility and ultralow magnetostriction. These wires reveal a robust hysteretic response, upon and after the application of external stress obtained in extreme levels of bending, reaching 5 mm of curvature radius. Longitudinal magneto-optic Kerr Effect measurements on magnetic thin films and arrays of nanowires and detailed analysis reveal a consistent reduction in magnetostriction constant moving from 3D bulk to 2D films and down to 1D nanowires, with almost one order of magnitude of reduction per step of reduced dimensionality. Our micromagnetic simulations attest observed results while suggesting that such changes originate from the reduction in size and dimensionality. The nanostructuring makes the flexible magnetic nanowires exhibit an enhanced elastic limit compared to thin films of cobalt and standard literature values for bulk cobalt. Furthermore, our samples display remarkable endurance that after nearly 100 bending cycles, the nanowires retain the original magnetic properties. The high mechanical stability of the magnetic properties and resilience opens new avenues of probing and exploiting strain-induced effects on interfaces and novel spacers in planar spintronic applications while being useful for creating advanced flexible applications in spintronics, magnonics, and magneto-plasmonics.

## Methods

### Sample preparation

Cobalt thin film and nanowire samples were prepared on flexible polyethylene naphthalate substrate obtained from DuPont. Nanowire samples were prepared by bilayer resist e-beam lithography that provides undercut for ease in liftoff. First, e-beam lithography was performed to make patterned stripes of nanowires with an aspect ratio of 10. Subsequently, a metallic film consisting of a layer of 3 nm of titanium, 30 nm of cobalt, and finally, a layer of 4 nm of aluminum to protect the cobalt layer from oxidation was deposited by e-beam evaporation. Finally, liftoff in hot acetone was performed to obtain the arrays of nanowires on a substrate ~ 1 cm x 1 cm.

### Magnetic characterization

Magnetic characterization was done using a home-built longitudinal magneto-optical Kerr effect (L-MOKE) setup. The MOKE setup was equipped with a red laser (wavelength 660 nm) modulated by a wave generator at 13 kHz and a lock-in amplifier SR 830 that allows collecting the signal from a Si photodetector with an amplitude of few tens of nA. To perform the stress-dependent magnetization measurements, we fabricated 3D printed sample holders having several radii of curvatures. The samples were placed on the curved base, and an upper frame-holder with an internal square window kept them in place. The frame ensures a uniform radius of curvature for the sample and allows the laser to hit the central area of the sample. To induce a controlled amount of stress, the samples were inserted into different sample holders, each with a specific curvature. To increase the signal-to-noise ratio, several hundred curves were measured and averaged to obtain the final M(H) curves. Besides, for each bending curvature, several hysteresis loops were measured for the same applied stress focusing the laser spot on different regions of the pattern, and the values of the parameters, such as the coercivity, were averaged to consider any possible minor variations of effective local stress applied in different positions of the samples.

### Micromagnetic simulations

Micromagnetic simulations were performed with MuMax$^3$, a GPU accelerated simulation software [28]. Commonly reported values of saturation magnetization $M_S$ = 1.4 MA/m and exchange stiffness constant $A_{ex}$ = 17.5 pJ/m of Co[46] were used to obtain results consistent with the experimental data. Periodic boundary conditions were applied repeating the simulation volume 10 times in X and Y directions to avoid limited size effects [47]. To accurately reproduce the sample, the system was discretized in cubic cells of ~ 2.7 nm edge, being smaller than the exchange length $l_{ex} = (2A/\mu_0 M_S^2)^{1/2}$ = 3.4 nm. To implement a random anisotropy model, the sample was divided into grains with an average grain size of 10 nm using a Voronoi tessellation.

**Acknowledgments:** The authors acknowledge the VR starting Grant (No. 2016-03278) from the Swedish Research Council, funding from the Olle Engkvist Foundation, and the Carl-Tryggers Foundation. The authors thank Olof Karis, Tapati Sarkar, and Jean-Francois Dayen for helpful feedback.

# Supporting Information

**Flexible ferromagnetic nanowires with ultralow magnetostriction**

Giuseppe Muscas[1,3], Petra E. Jönsson[1], I. G. Serrano[1], Örjan Vallin[2], and M. Venkata Kamalakar[1]*

[1]Department of Physics and Astronomy, Uppsala University, Box 516, SE-751 20 Uppsala, Sweden
[2]Department of Engineering Sciences, Uppsala University, Box 534, SE-751 21 Uppsala, Sweden
[3]Department of Physics, University of Cagliari, Cittadella Universitaria di Monserrato, S.P. 8 Km 0.700 (Monserrato-Sestu), I-09042 Monserrato CA, Italy

*Email: venkata.mutta@physics.uu.se

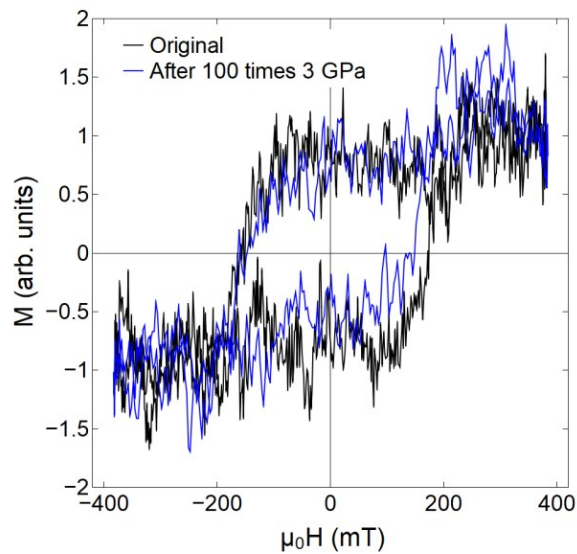

Figure S1. M(H) loop measured along the easy axis of Co nanowires with size 1.0 µm X 100 nm. The curve has been measured for the as-prepared original sample (black curve) and after a stress test (blue curve) after the application of 3 GPa of tensile stress along the width of the wires 100 times.